\documentclass[draft]{agujournal2019}
\usepackage{url} 
\usepackage[inline]{trackchanges} 
\usepackage{soul}
\usepackage{graphicx}
\usepackage{subcaption}

\draftfalse

\journalname{JGR: Atmospheres}

\begin{document}

%
%


\title{Underlying Physical Mechanisms in Upward Positive Flashes}

%
%




\authors{T. Oregel-Chaumont\affil{1}, A. \v{S}unjerga\affil{2}, J. Kasparian\affil{3}, M. Rubinstein\affil{4}, and F. Rachidi\affil{1}}

\affiliation{1}{Swiss Federal Institute of Technology (EPFL)}
\affiliation{2}{University of Split}
\affiliation{3}{University of Geneva}
\affiliation{4}{University of Applied Sciences and Arts Western Switzerland}





\correspondingauthor{Toma Oregel-Chaumont}{toma.chaumont@epfl.ch}




\begin{keypoints}
\item We present the first observation of a mixed-mode of charge transfer during an upward positive flash.
\item We describe a physical mechanism underpinning Type 1 and Type 2 upward positive flashes. 
\item 
The findings challenge existing classifications for mixed mode and M-component--type, confirming they form a single mode of charge transfer.

\end{keypoints}

%
%

%
%


\begin{abstract}

This study presents the first observation of a mixed mode of charge transfer during an upward positive flash, which was initiated from the S\"antis Tower in Switzerland.
High-speed camera footage, along with current and electric field measurements, revealed a downward-propagating recoil leader connecting to the grounded current-carrying plasma channel at a junction height of $<$1~km above the tip of the tower. 
This event triggered the ``return stroke''-like main pulse associated with Type 1 upward positive flashes, leading us to propose a mixed mode of charge transfer (normally observed in upward \textit{negative} flashes) as the physical mechanism at play.
Furthermore, the observed `Main pulse' shared characteristics with both mixed-mode and M-component--type initial continuous current (ICC) pulses, challenging existing classification criteria, and supporting the notion of a unique mode of charge transfer with a range of junction length-dependent pulse characteristics, as opposed to two distinct modes.
The recoil leader itself was accompanied by a sequence of fast electric field pulses indicative of step-like propagation, also an observational first.
These findings contribute to improving our understanding of the mechanisms of charge transfer in upward lightning flashes. 
\end{abstract}

\section*{Plain Language Summary}
Despite its ubiquitous nature, lightning is a complex phenomenon that scientists are still working to understand fully. 
In addition to the more common downward, cloud-to-ground strikes, lightning can also propagate \textit{upwards}, typically from the tips of tall towers, wind turbines, or airplanes, and can exchange both positive \textit{and/or} negative charge in the process. 
This study discusses a particular upward positive flash that occurred at the S\"antis Tower in Switzerland.
Using high-speed camera footage, electric field sensors, and a current measurement system, 
the analysis presented in the paper allowed us to propose a physical mechanism describing the two different categories of upward positive flashes.
%
%

%


%
%
%
%

\section{Introduction}
\label{sec:intro}

The physics of upward lightning is poorly understood compared to the more common downward lightning, especially when it comes to the even rarer positive flashes.
As such, advancing knowledge often relies on building upon better-studied phenomena.
Upward lightning initiated from tall structures typically includes an initial stage (IS), which may be followed by one or more sequences of downward-propagating dart leaders and corresponding upward return strokes.
These sequences are analogous to those associated with subsequent return strokes observed in natural downward lightning.
The IS itself comprises the upward leader and the initial continuous current (ICC).
Often superimposed on the slowly-varying ICC, are current pulses appropriately known as ICC pulses, which have been observed in both upward lightning from tall towers and rocket-triggered lightning.

\citeA{miki_initial_2005} investigated ICC pulses characterized by short rise times.
Their findings indicate that such pulses are associated with low-altitude upward branching in object-initiated lightning.
Specifically, when two upward branches are present—one luminous and the other decayed—it is possible for a continuous current to flow in the luminous branch, while a dart leader develops in the decayed branch.
These branches typically share a common channel section leading to the top of the strike object.
\citeA{miki_initial_2005} interpreted the resulting ICC pulses as manifestations of the leader/return-stroke mode of charge transfer to ground within the newly-illuminated (decayed) branch, a concept originally proposed by \citeA{rakov_lightning_2003}.

In a related study, \citeA{winn_luminous_2012} analyzed high-speed video recordings of luminous pulses—corresponding to ICC pulses—during the initial stage of a rocket-triggered flash in New Mexico (elevation $\sim$3~km).
They proposed a unified explanation for ICC pulses: the merger of a dart leader traveling through a relatively dark branch into an already illuminated one.

\citeA{warner_upward_2012} presented a similar picture, further stating that the dart leader attached to the actively luminous main channel that was already connected to the top of a tall object. 
The dart leader was moreover observed to be a bidirectional recoil leader.
\citeA{yoshida_initial_2012} observed, in agreement with the above, that the ICC pulses can indeed be the result of the connection of recoil leaders to the main active channel, but they also observed, using VHF images, that ICC pulses can also be associated with stepped leaders in the cloud that attach to a current-carrying channel connected to ground in triggered lightning.

Examining simultaneous measurements of currents, electric fields, and high-speed video images associated with upward flashes initiated from the Gaisberg Tower in Austria, \citeA{zhou_study_2015} showed that two parallel channels—one pre-existing and the other newly formed or reactivated—participate in the charge transfer process, sharing a common channel segment between their junction point and the strike object.
They proposed the term ``mixed mode of charge transfer to ground'' to describe this scenario.
In such cases, where the junction point is located at a relatively low altitude, the classical M-component mode cannot occur, despite the presence of a conductive path to ground, which is typically a defining characteristic of the M-component mode.

\citeA{he_characteristics_2020} analyzed 94 pulses that occurred during two upward negative flashes recorded at the S\"antis Tower.
The obtained results supported the assumption that the mode of charge transfer to the ground giving rise to mixed-mode pulses is similar to that of return strokes.

In view of the above, unlike downward lightning flashes—where three primary modes of charge transfer are typically identified \cite{rakov_lightning_2003} to describe how charge is lowered in cloud-to-ground flashes (leader--return-stroke sequences, continuing currents, and M-components), upward lightning flashes involve a broader set of charge transfer modes:
1. Initial Continuous Current (ICC);
2. Leader--return-stroke sequences;
3. Continuing Currents;
4. M-Components; and,
5. ICC pulses (namely pulses superimposed on the ICC). 
Modes 2 through 4 are analogous to those observed in downward lightning.
The ICC pulses can be further categorized into two distinct types:
\begin{itemize}
    \item 5a. M-Component--type or slow pulses. These pulses are associated with the reactivation of a decayed branch or the connection of a newly created channel to the ICC-carrying channel at relatively large junction heights;
    \item 5b. Mixed Modes, as defined by \citeA{zhou_mixed_2011, zhou_study_2015} or fast pulses. Mixed mode pulses are associated with the reactivation of a decayed branch or the connection of a newly created channel to the ICC-carrying channel at relatively small junction heights.
\end{itemize}
Mixed mode pulses are associated with processes similar to leader/return-stroke events occurring in decayed or newly-created branches of the plasma channel connecting to the grounded, current-carrying channel, with junction points below the cloud base (height $<$1~km AGL, as opposed to M-Component--type ICC pulses with junction heights above 1~km, and characteristics more similar to classical M-components) \cite[]{zhou_study_2015}.

In addition to this junction point height cutoff they defined at $\sim$1~km, three other criteria have also been proposed in the literature to distinguish M-component--type ICC pulses and mixed-mode pulses \cite{li_review_2023}: 
the lag between the channel-base current pulse and its associated radiated electric field pulse \cite{zhou_study_2015}, the current pulse rise time \cite{flache_initialstage_2008}, and the pulse waveform asymmetry \cite{he_analysis_2018}. [Note that the time lag was \citeA{zhou_study_2015}'s primary distinguishing criterion, and they identified the attachment-point height as equivalent; see their Figure 15.]


Heretofore, mixed-mode ICC pulses have only been observed in upward negative flashes \cite{zhou_mixed_2011, he_characteristics_2020}.
Upward \textit{positive} flashes, a much rarer phenomenon, are also characterized by an ICC phase punctuated by pulses, though these are due to the stepping of the upward negative leader \protect\cite{qie_propagation_2019}. 
As further discussed in Section~\ref{sec:dis}, two types of positive flashes have been identified in the literature: Type 1 flashes, which exhibit a large unipolar return stroke–like current pulse following the upward negative stepped-leader phase, and Type 2 flashes, which lack such a large pulse. 
This classification is based solely on current waveform characteristics. 
Thanks to simultaneous measurements of current, electric field and high-speed camera images obtained at the S\"antis Tower, we can now propose a physical mechanism distinguishing Type 1 and Type 2 upward positive flashes.
Herein, we report, to the best of our knowledge, the first observation of the mixed mode of charge transfer during an upward positive flash, which was initiated from the S\"antis Tower in Switzerland. 
Via an analysis of the simultaneous records provided, we propose this `mixed mode', triggered by a downward-connecting recoil leader, as the physical mechanism responsible for the `main pulse' observed in Type 1 upward positive flashes.


\section{Materials \& Methods / Experimental Setup}


\begin{figure}
    \centering
    \includegraphics[width=0.9\textwidth]{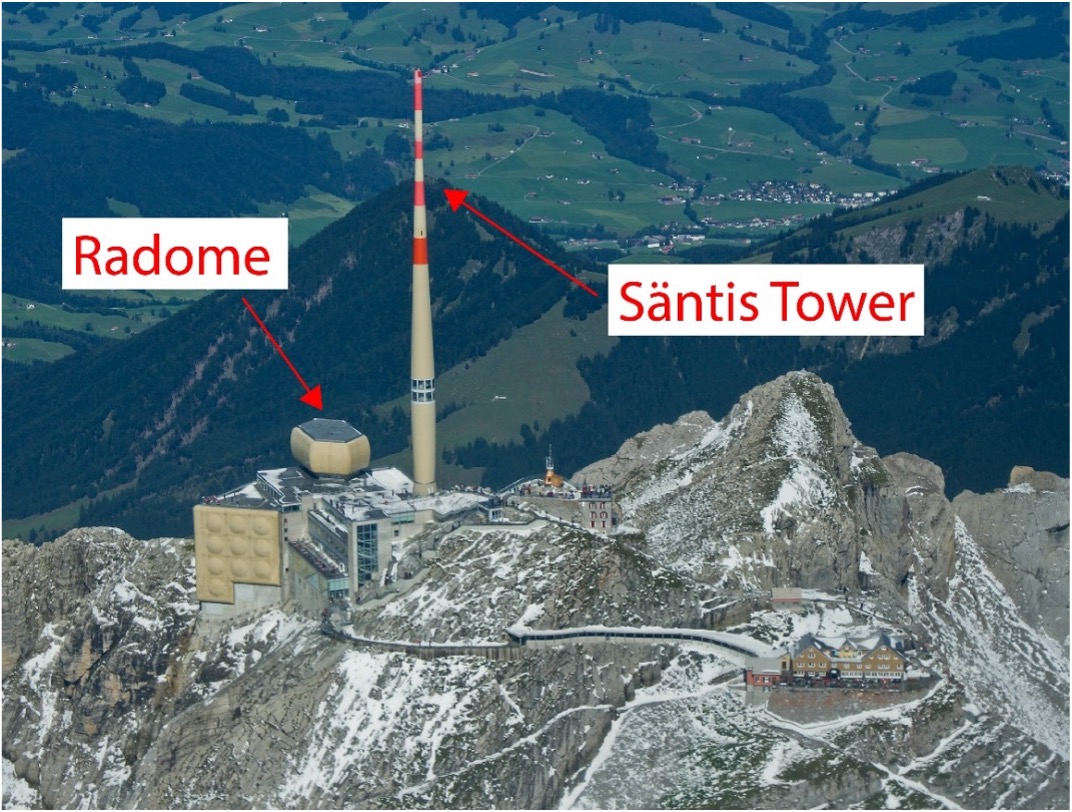}
    \caption{Photo of the S\"antis peak, with arrows indicating the Radome, which houses an electric field probe, and the Tower, where the current sensors are located. Image reproduced from \protect\citeA{rachidi_santis_2022} (Fig. 2) with permission.}
    \label{fig:SantisPic}
\end{figure}


The Mt. S\"antis Lightning Research Facility, shown in Figure~\ref{fig:SantisPic}, is situated at 2502 m ASL in the Appenzell Alps of north-eastern Switzerland, and experiences $>$100 direct lightning strikes per year to its 124 meter-tall tower, which is equipped with a comprehensive current measurement system consisting of
Rogowski coil and B-dot sensor pairs at two different heights: 24 and 82 meters above ground level (AGL). 
The nearby Radome (20 m from the Tower base) houses an E-field sensor with a sampling rate of 20 MHz. 
This M\'elop\'ee fast E-field probe has a frequency range of 1 kHz to 150 MHz, a time constant of 160 $\mu$s, and is described in more detail in \citeA{sunjerga_santis_2021}.
Five kilometers away, atop Mt. Kr\"onberg (1663 m ASL), is a high-speed camera (HSC) operating at 24,000 fps, with an exposure time of $\sim 41~\mu$s and a resolution of 512$\times$512 pixels.
Electric field measurements are also taken 15 km away by a flat-plate antenna with line-of-site in Herisau, Switzerland, which has a frequency range of 40 Hz to 40 MHz and a time constant of 8~ms.

Additionally, during the Summer of 2021, when the flash discussed below occurred, a second high-speed camera operating at 11,500 fps with $512\times512$ pixel resolution was installed in Schw\"agalp, at the base of Mt. S\"antis (line-of-sight distance of $\sim$2~km from the Tower), and was used in conjunction with the Kr\"onberg HSC to estimate the 3D velocities of the upward negative leader. 

The Radome E-field signal is synchronized with the tower current signal by aligning the time of the first E-field ``step'' with the time of the first current pulse, i.e., the first significant deviation from zero in both waveforms, associated with the onset of the upward negative leader and the ICC, respectively.
The high-speed cameras and ``far'' E-field data are synchronized with the rest by GPS timestamp if the antennae are functional at the time of the flash. 
If not, manual synchronization can be carried out via waveform matching. 
More detailed information on the S\"antis measurement system can be found in \citeA{rachidi_santis_2022}.

All computational data analysis and presentation were carried out using MATLAB and the Python programming language, in particular the NumPy, SciPy, and Matplotlib libraries.


\section{Results / Observations}

%
\begin{figure}
    \centering
    \begin{subfigure}{0.56\textwidth}
        \centering
        \includegraphics[width=\textwidth]{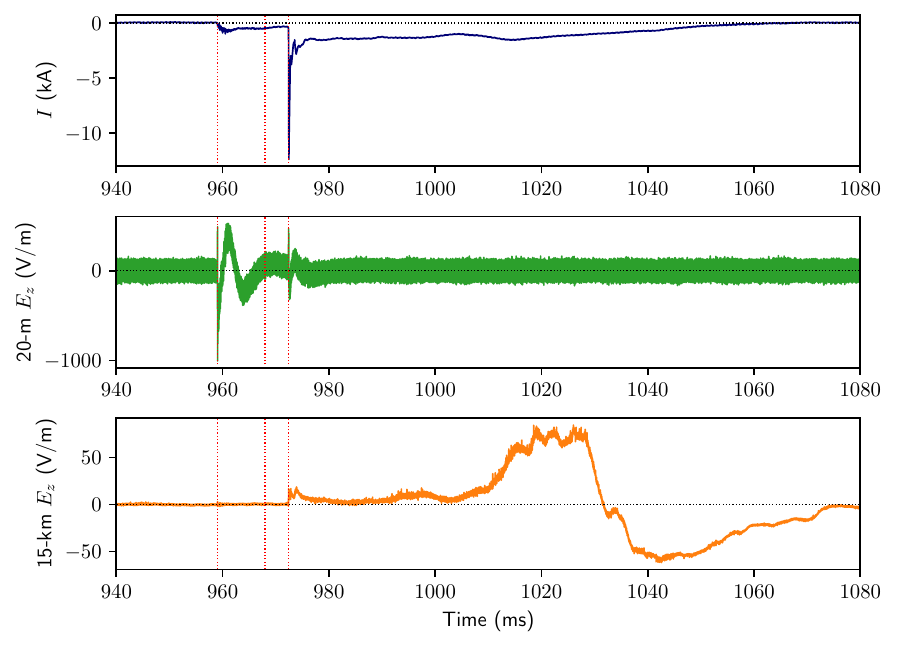}
    \end{subfigure}
    \hfill
    \begin{subfigure}{0.42\textwidth}
        \centering
        \includegraphics[width=\textwidth]{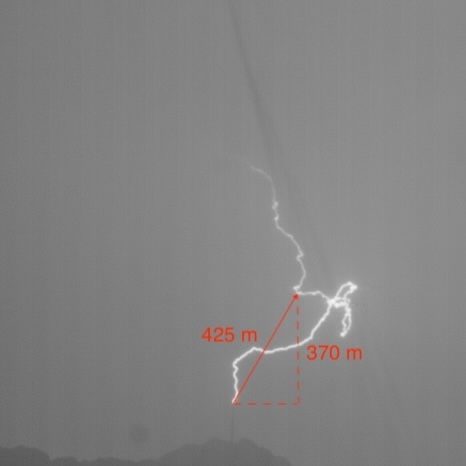}
    \end{subfigure}
    \caption{Waveforms (current, near and ``far'' E-field) of the observed upward positive flash that occurred on July 24, 2021 at 16:24:03 UTC. Top left waveform: tower current. A 100~kHz low-pass filter was applied to remove intermittent noise (though the peak current and rise time were determined from the raw data), and we chose the sign convention of a negative current corresponding to a positive charge transfer from cloud to ground. Middle and bottom left waveforms: measured vertical electric field at 20~m (Radome) and 15~km (Herisau), respectively, using the physics sign convention. The Herisau E-field was manually synchronized with the other two waveforms, and therefore accounts for the $\sim$50~$\mu$s time delay. Time is measured from the beginning of the recording ($\sim$1~s before the current peak). The vertical red lines highlight the three phases of interest (from left to right: upward negative leader, recoil leader-channel creation, and main pulse) further analyzed in figures~\ref{fig:UNL}-\ref{fig:RL-MP}. The image on the right is an HSC-frame integration over the duration of the flash. Superposed red lines identify the 2D distances discussed in-text.} 
    \label{fig:wholeflash} 
\end{figure}

\begin{figure}
    \centering
    \begin{subfigure}[t]{0.32\textwidth}
        \includegraphics[width=\textwidth]{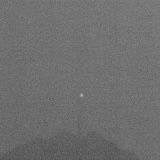}
        \includegraphics[width=\textwidth]{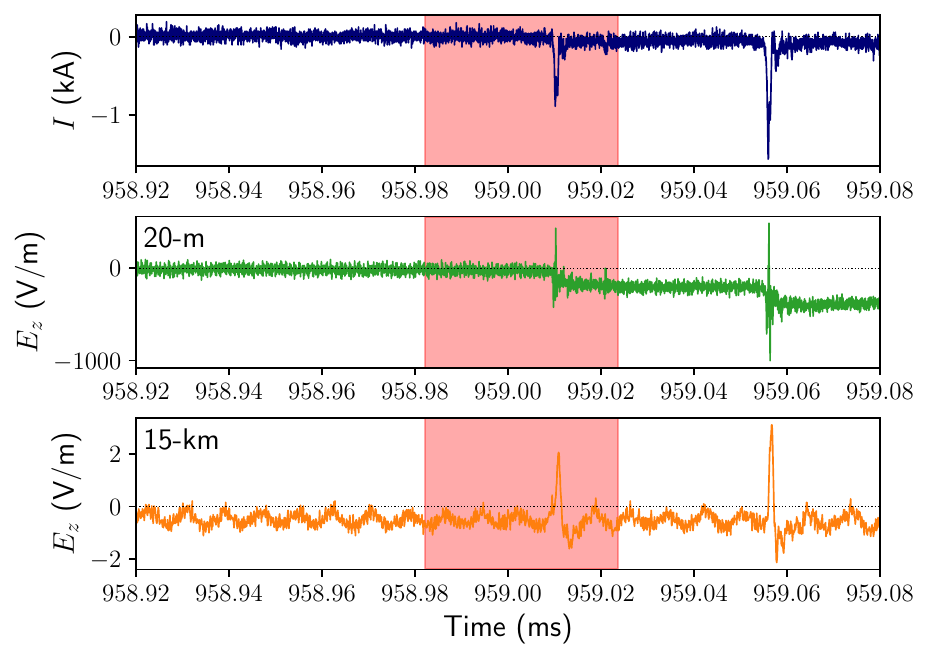}
        \subcaption*{Frame 654}
    \end{subfigure}
    \begin{subfigure}[t]{0.32\textwidth}
        \includegraphics[width=\textwidth]{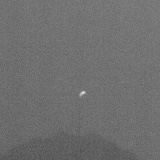}
        \includegraphics[width=\textwidth]{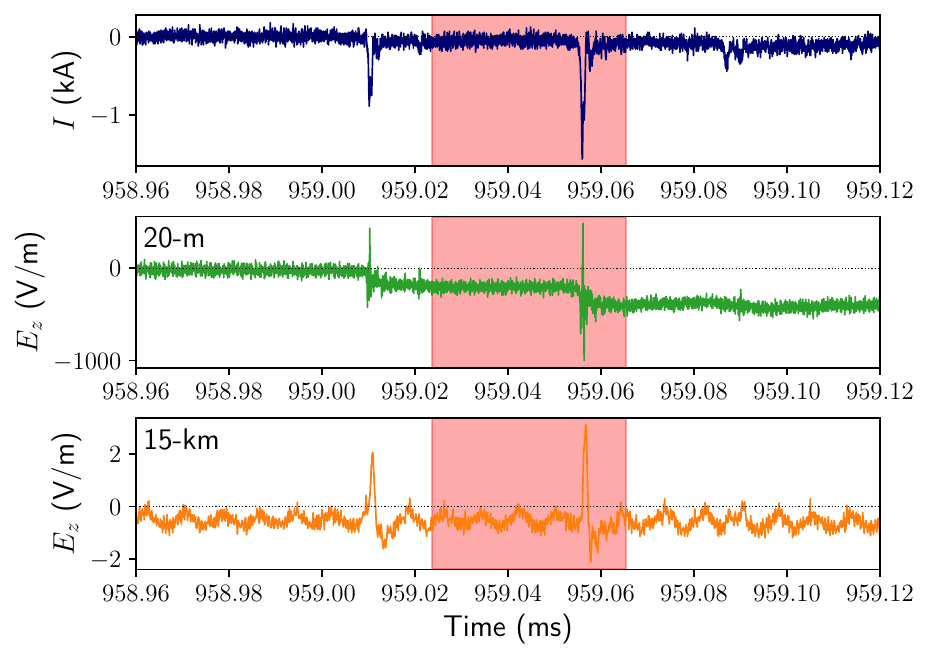}
        \subcaption*{Frame 655}
    \end{subfigure}
    \begin{subfigure}[t]{0.32\textwidth}
        \includegraphics[width=\textwidth]{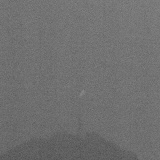}
        \includegraphics[width=\textwidth]{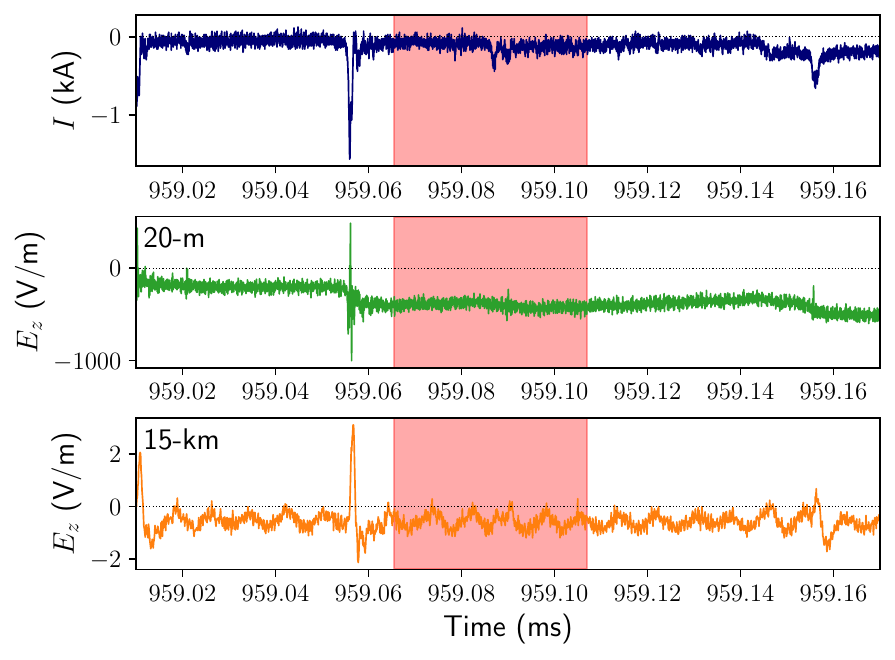}
        \subcaption*{Frame 656}
    \end{subfigure}
    \begin{subfigure}[t]{0.32\textwidth}
        \includegraphics[width=\textwidth]{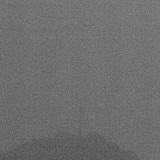}
        \includegraphics[width=\textwidth]{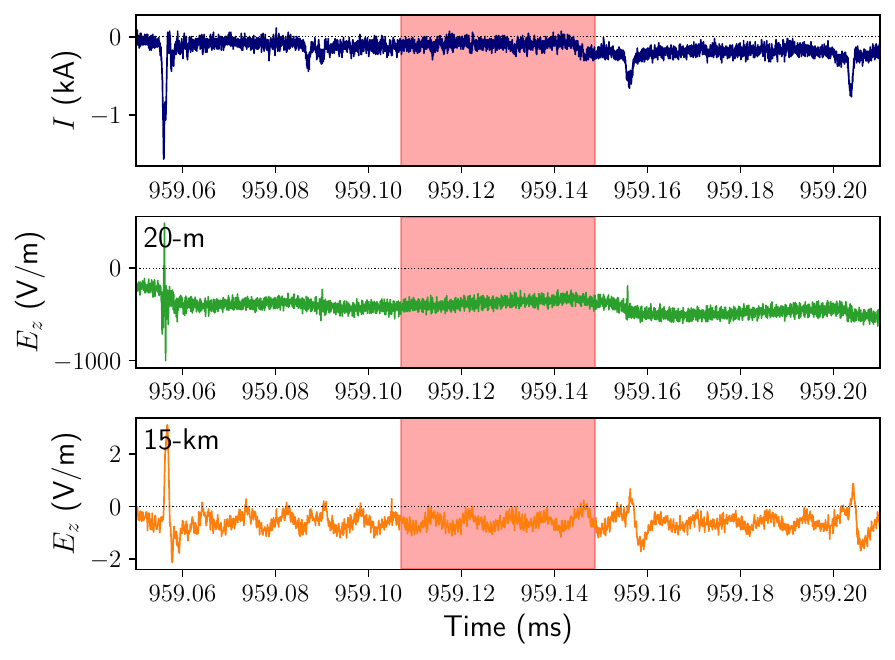}
        \subcaption*{Frame 657}
    \end{subfigure}
    \begin{subfigure}[t]{0.32\textwidth}
        \includegraphics[width=\textwidth]{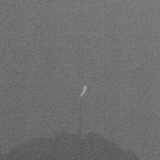}
        \includegraphics[width=\textwidth]{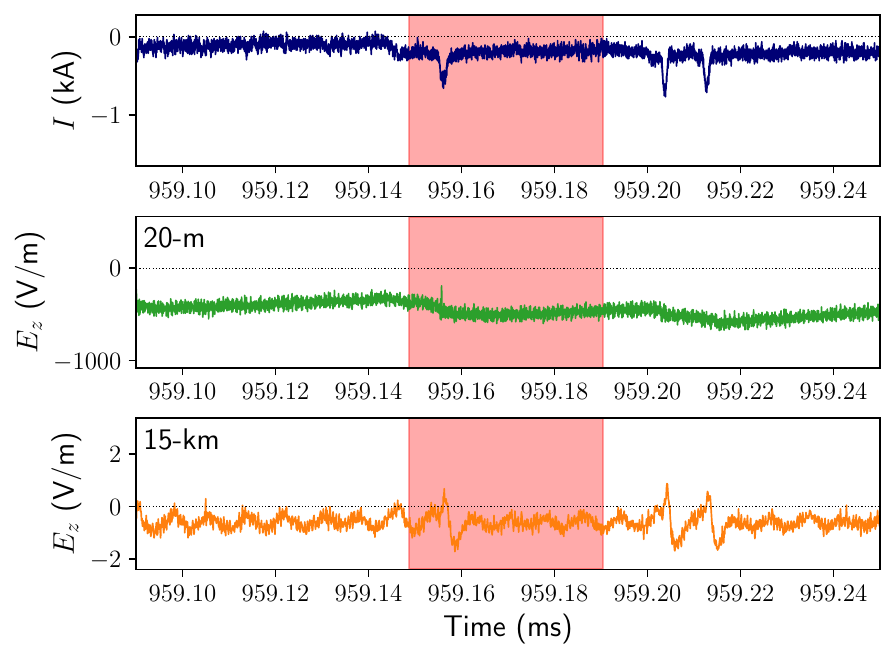}
        \subcaption*{Frame 658}
    \end{subfigure}
    \begin{subfigure}[t]{0.32\textwidth}
        \includegraphics[width=\textwidth]{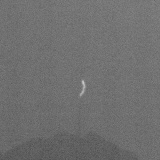}
        \includegraphics[width=\textwidth]{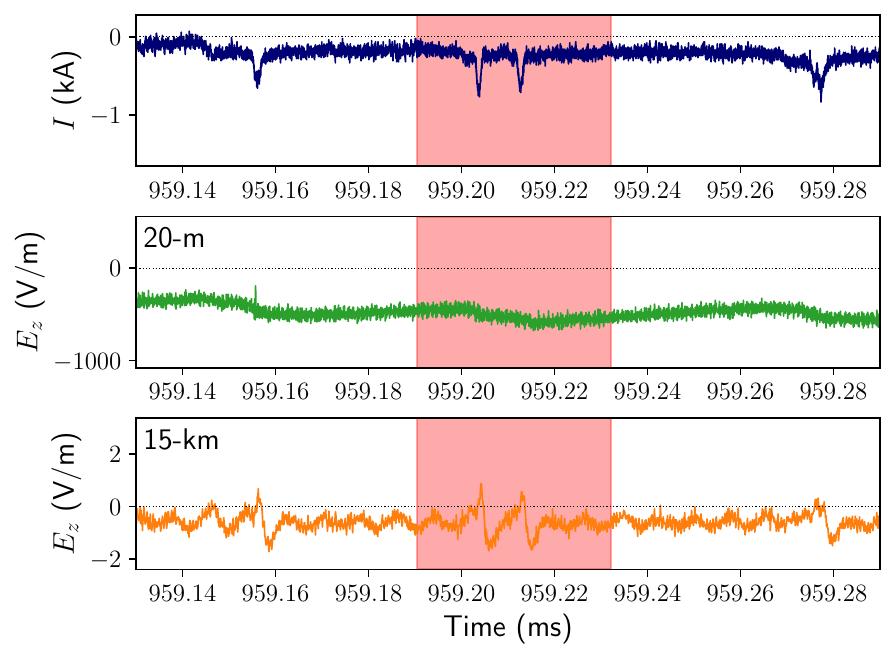}
        \subcaption*{Frame 659}
    \end{subfigure}
    \caption{The upward negative stepped leader phase of the flash. The first six HSC frames are shown above. Their approximate temporal widths (42~$\mu$s) are highlighted by the red-shaded regions in the waveforms below, which depict, from top to bottom, the tower current (blue), 20-m E-field (green), and 15-km E-field (orange), as before. See \protect\citeA{oregel-chaumont_direct_2024} for a detailed discussion of this phase, and Figure~\ref{fig:wholeflash} 
    for a zoomed-out view of the waveforms.}
    \label{fig:UNL}
\end{figure}

The upward positive flash observed and analyzed in this study initiated from the S\"antis Tower at 16:24:03 UTC on July 24th, 2021.
The integrated high-speed camera frames, as well as the current and electric field waveforms, for the whole flash are shown in Figure~\ref{fig:wholeflash}.
In the right-hand panel of Figure~\ref{fig:wholeflash}, one can discern the S\"antis tower, from which the flash initiated, at the base of the rather tortuous plasma channel. [The faint black streak running diagonally across the integrated stills is formed by raindrops beading on the camera’s protective windowpane.]
A superimposed red arrow indicates the 2D distance from the tower tip to the junction point, where a downward-propagating leader connected to the conducting portion of the plasma channel, after retracing a decayed channel.
The tower current waveform (top left panel) features the ICC associated with the upward negative leader, which lasted for 13.45~ms before being punctuated by a ``return stroke''-like main pulse, followed by a couple minor M-component--type pulses, then a gradually-decaying current.
The close electric field (Radome, 20~m) waveform in the middle left panel features large changes associated with the upward negative leader, whose stepping also produced small pulses in the 15~km electric field waveform (lower left panel), though these are not visible at this scale; see Figure~\ref{fig:UNL}.
The ``return stroke''-like main pulse is also present in the E-field waveforms, albeit to a lesser degree in the Radome.
Note that the large changes midway through the 15-km Herisau E-field waveform are likely due to unrelated nearby lightning activity.
Highlighted by dotted vertical red lines in all three Figure~\ref{fig:wholeflash} waveforms are the three phases of greatest interest: (a) initiation of the upward negative leader, (b) creation of the plasma channel branch along which the recoil leader will later propagate, and (c) the main pulse following the connection of the recoil leader to the current carrying channel.


%


Figure~\ref{fig:UNL} provides a close-up of phase (a).
We observe the characteristic ICC pulses and electric field changes associated with the creation and propagation of an upward negative stepped leader.
These are attenuated over time due to signal propagation effects, but the baseline current and near E-field continue to increase as the plasma channel extends (this static component is lost in the far-field, though the radiated ``steps'' remain distinguishable).


%
\begin{figure}
    \centering
    \begin{subfigure}[t]{0.32\textwidth}
        \includegraphics[width=\textwidth]{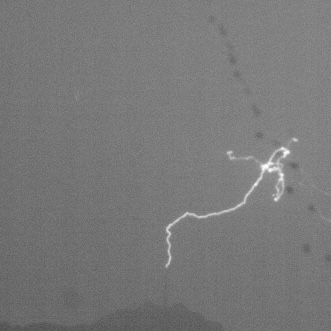}
        \includegraphics[width=\textwidth]{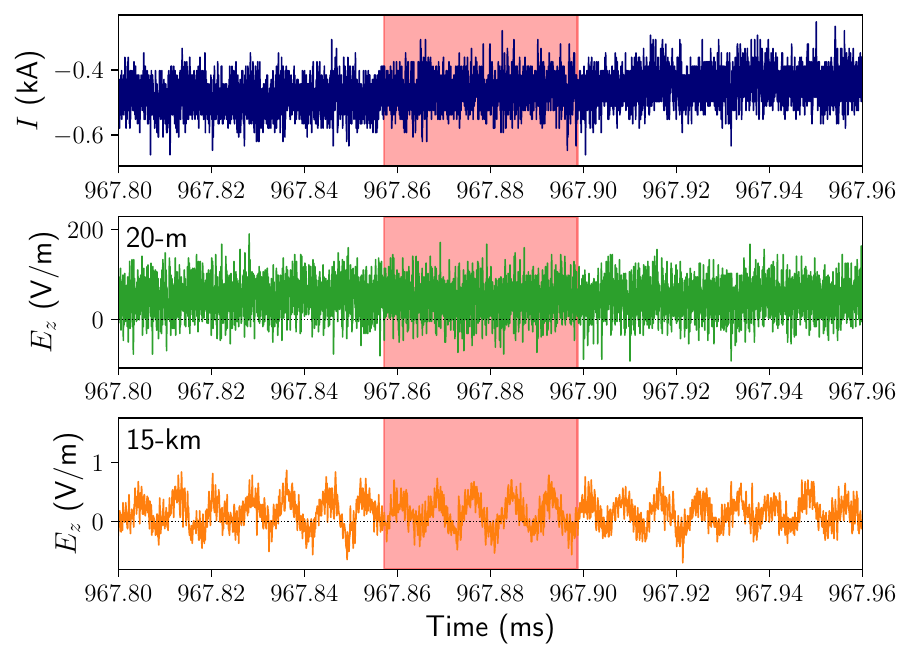}
        \subcaption*{Frame 866}
    \end{subfigure}
    \begin{subfigure}[t]{0.32\textwidth}
        \includegraphics[width=\textwidth]{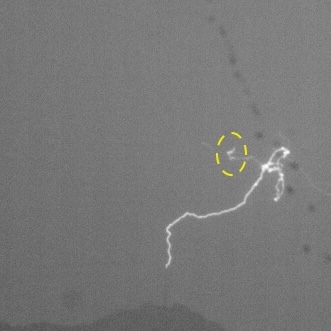}
        \includegraphics[width=\textwidth]{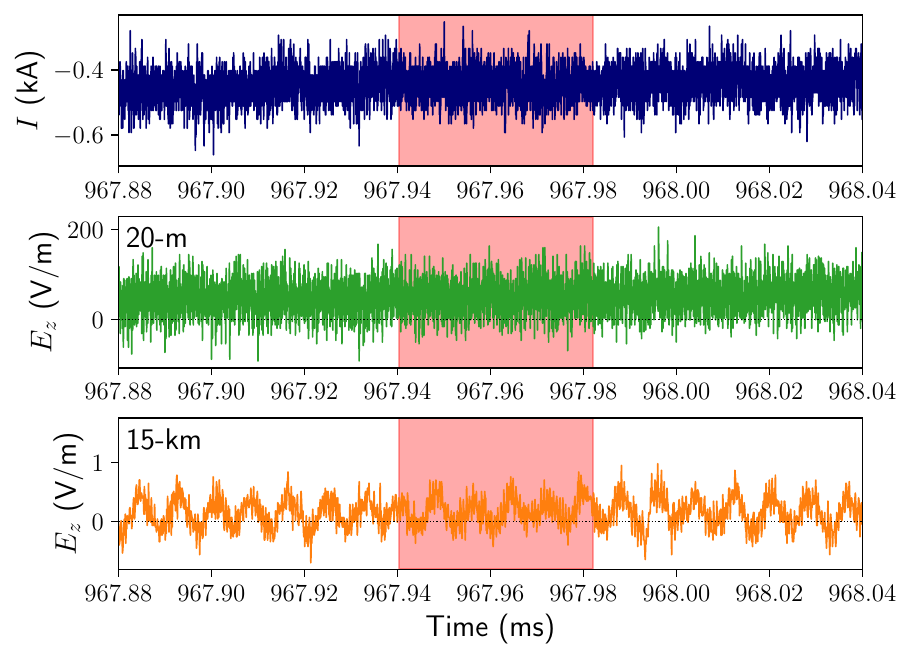}
        \subcaption*{Frame 868}
    \end{subfigure}
    \begin{subfigure}[t]{0.32\textwidth}
        \includegraphics[width=\textwidth]{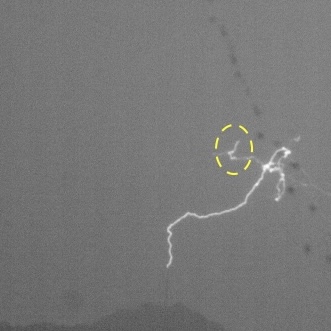}
        \includegraphics[width=\textwidth]{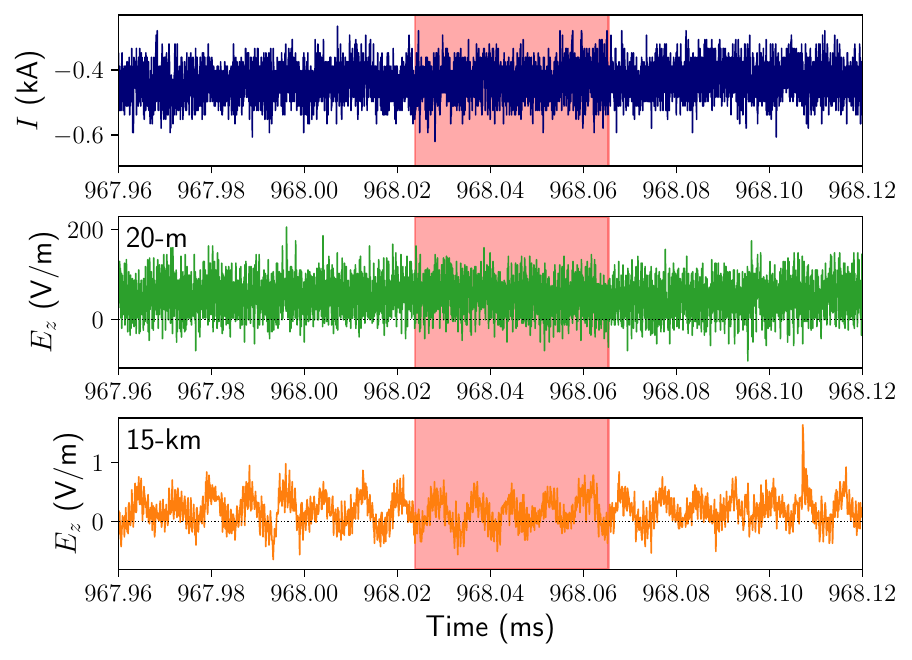}
        \subcaption*{Frame 870}
    \end{subfigure}
    \begin{subfigure}[t]{0.32\textwidth}
        \includegraphics[width=\textwidth]{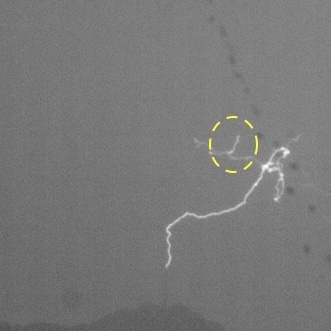}
        \includegraphics[width=\textwidth]{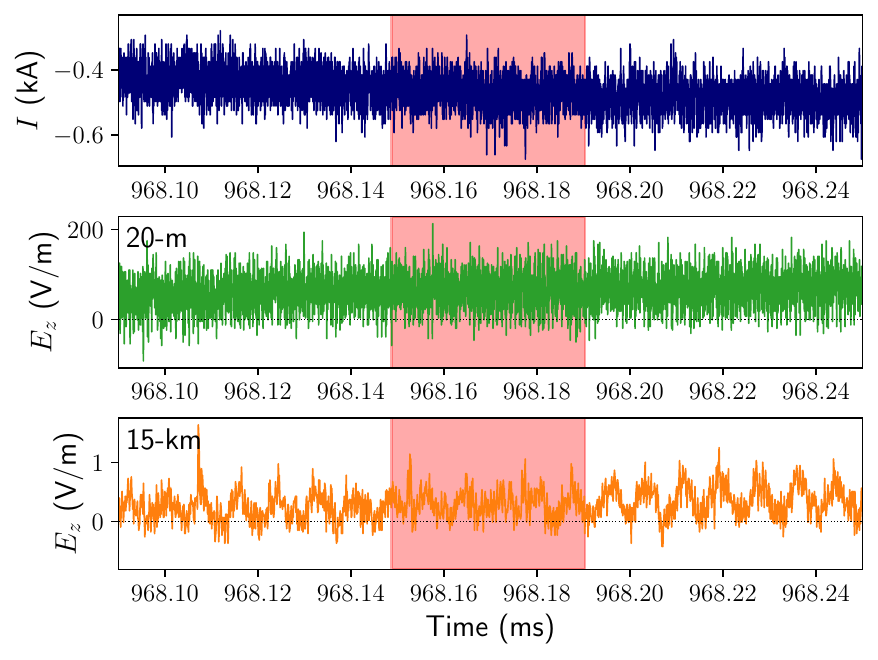}
        \subcaption*{Frame 873}
    \end{subfigure}
    \begin{subfigure}[t]{0.32\textwidth}
        \includegraphics[width=\textwidth]{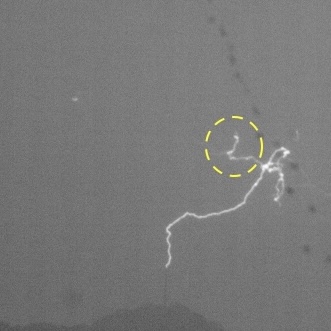}
        \includegraphics[width=\textwidth]{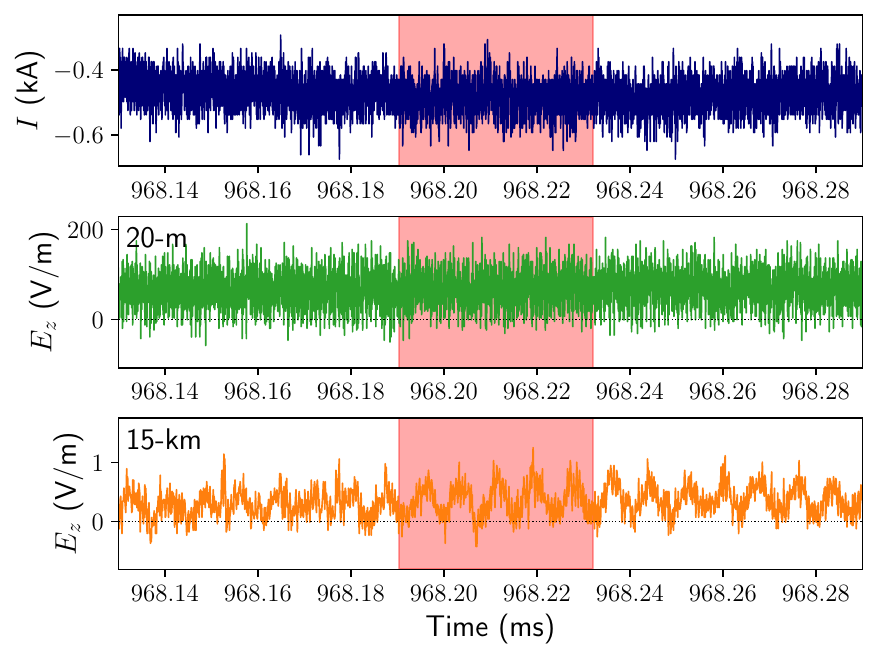}
        \subcaption*{Frame 874}
    \end{subfigure}
    \begin{subfigure}[t]{0.32\textwidth}
        \includegraphics[width=\textwidth]{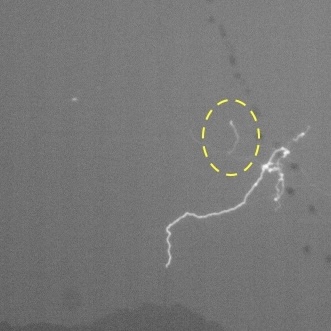}
        \includegraphics[width=\textwidth]{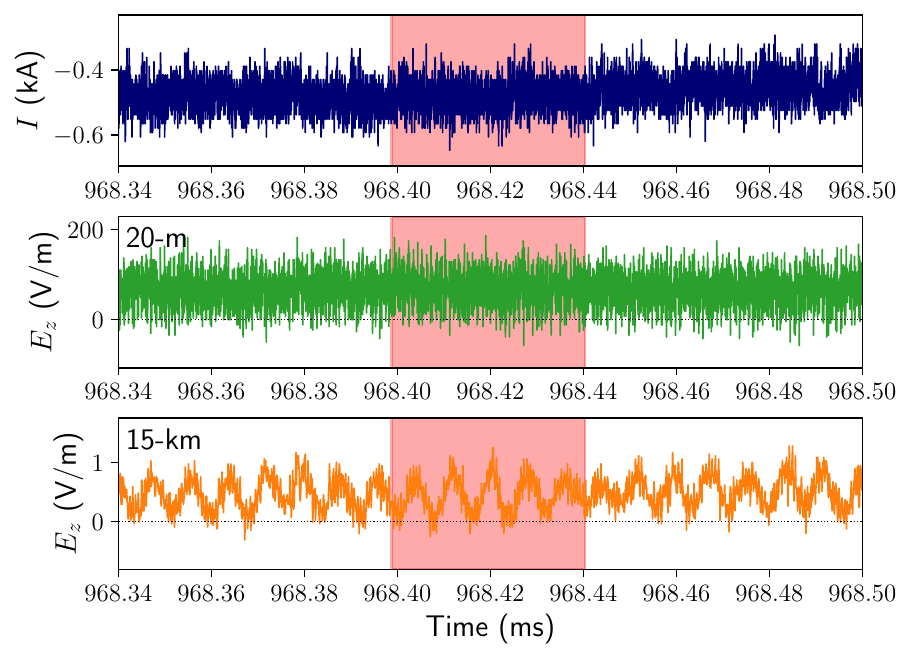}
        \subcaption*{Frame 879}
    \end{subfigure}
    \caption{The creation of the plasma channel (2D junction height of 495$\pm$5~m AGL) along which the recoil leader will later propagate (highlighted in yellow). See Figure~\ref{fig:branchHSC} for more frames and details. The waveforms are as described in Figure~\ref{fig:UNL}.}
    \label{fig:branch}
\end{figure}


Figure~\ref{fig:branch} provides a close-up of phase (b): the branch-creation process.
We see from the HSC frames how this new plasma channel extends from the junction point, reflected in the waveforms as a steady continuous current of $-0.50\pm0.05$~kA.
Yellow dashed ellipses highlight the lengthening branch in each frame.
The prior formation of this plasma channel provides evidence that the main pulse to come was preceded by specifically a recoil leader.
The absence of waveform pulses associated with the stepping process can have several explanations, and is likely a combination of them:
1. For the current, this may be due to further attenuation of the signal as explained above.
2. In the 20-m E-field, fast pulses become indistinguishable due to the dominance of the electrostatic field component.
3. In the 15-km E-field, pulse amplitude appears to decrease to the point of being lost in the noise ($\sim$1~V/m) after $\sim$2~ms. This decrease has been observed in the literature \cite{kolmasova_rapid_2025}, particularly when currents travel through channels with a strong horizontal component, as is the case here (the main channel starts propagating horizontally after $\sim$1~ms).
4. It has further been observed (Warner et al. 2012b) that the frequency of initial breakdown pulses in upward leaders can also decrease as time goes on, i.e., the leader transitions from propagating in a ``step-like'' manner to a more ``continuous'' one.
5. Though to date only observed near positive leaders, an excited ``floating'' channel could also explain the non-observation of waveform pulses \cite{montanya_start_2015, warner_observations_2016, yuan_development_2019}.


%
\begin{figure}
    \centering
    \begin{subfigure}{0.32\textwidth}
        \includegraphics[width=\textwidth]{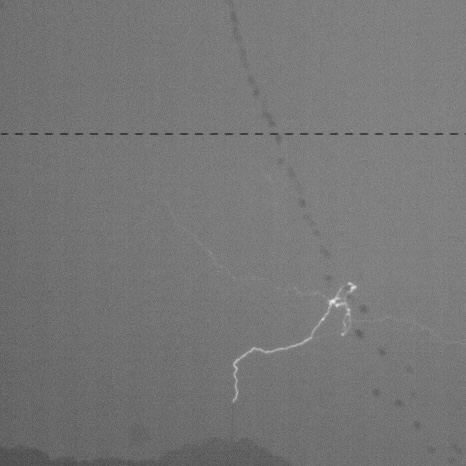}
        \includegraphics[width=\textwidth]{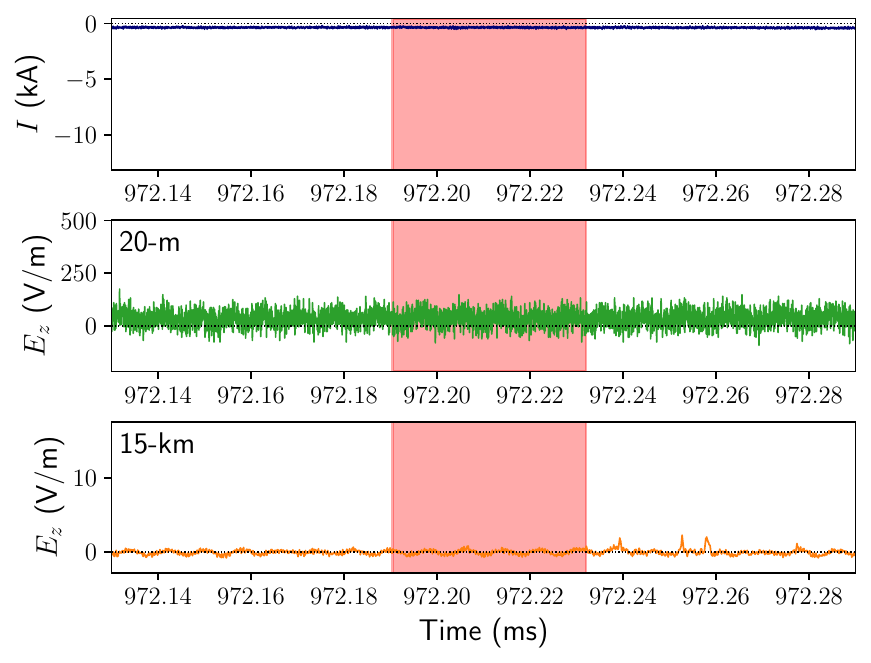}
        \subcaption*{Frame 971}
    \end{subfigure}
    \begin{subfigure}{0.32\textwidth}
        \includegraphics[width=\textwidth]{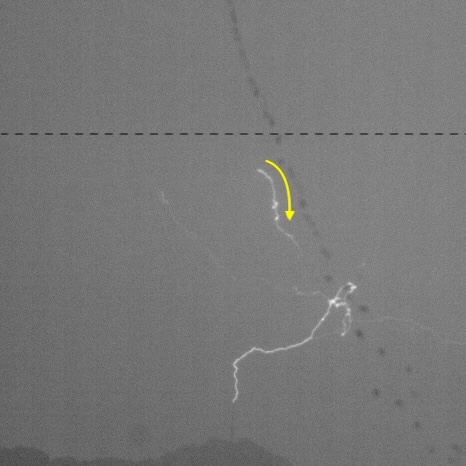}
        \includegraphics[width=\textwidth]{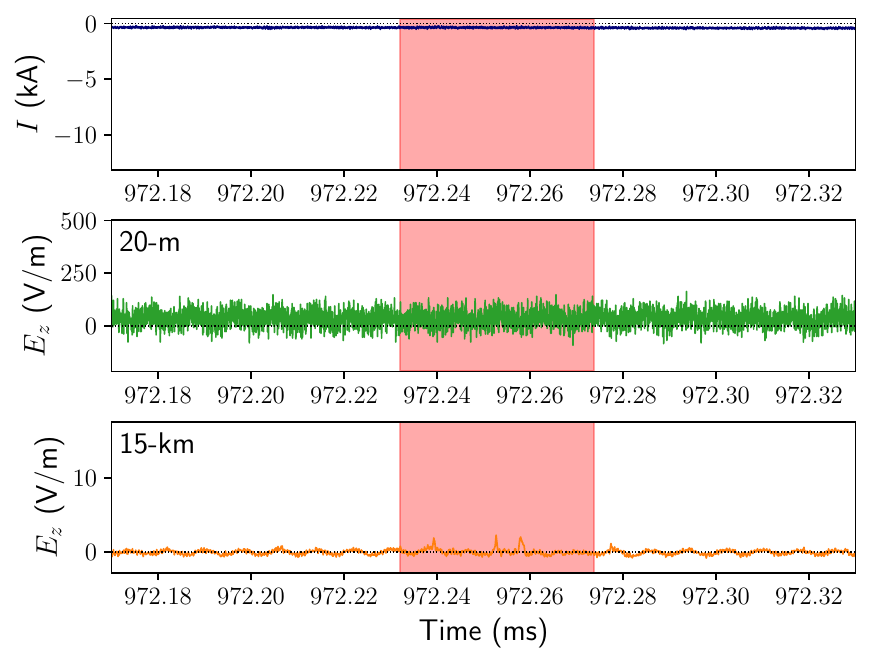}
        \subcaption*{Frame 972}
    \end{subfigure}
    \begin{subfigure}{0.32\textwidth}
        \includegraphics[width=\textwidth]{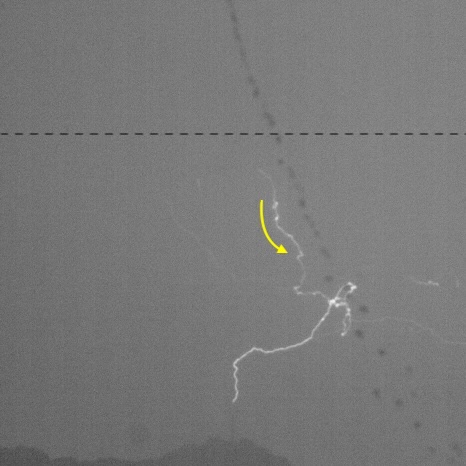}
        \includegraphics[width=\textwidth]{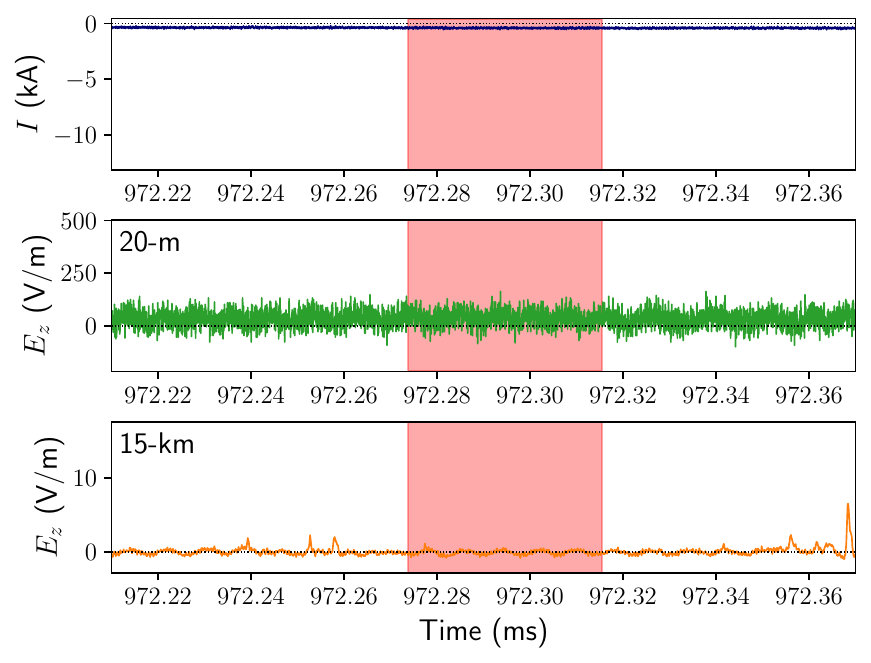}
        \subcaption*{Frame 973}
    \end{subfigure}
    \begin{subfigure}{0.32\textwidth}
        \includegraphics[width=\textwidth]{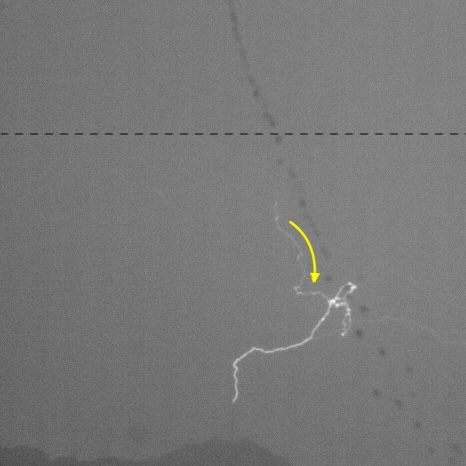}
        \includegraphics[width=\textwidth]{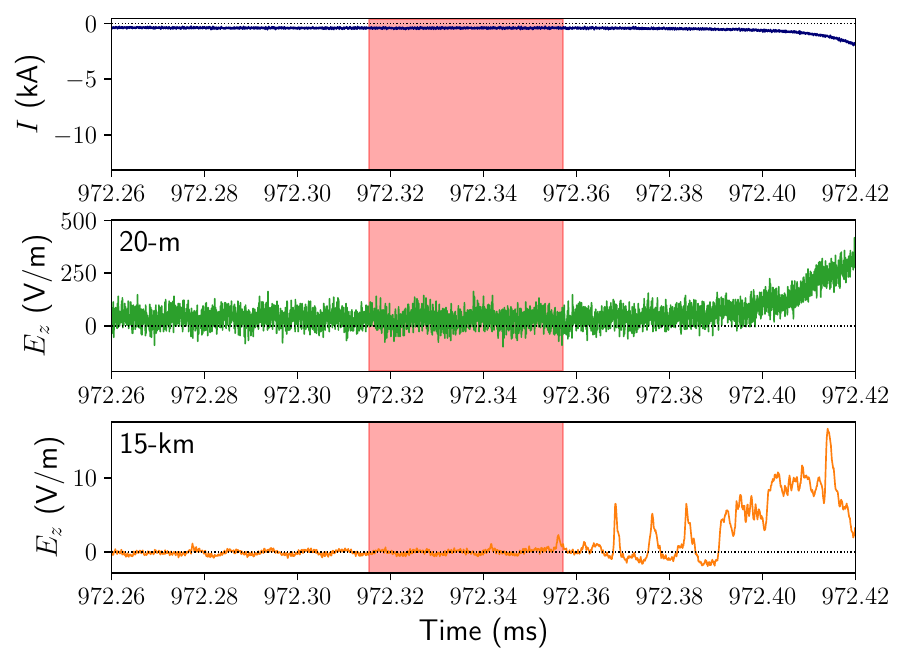}
        \subcaption*{Frame 974}
    \end{subfigure}
    \begin{subfigure}{0.32\textwidth}
        \includegraphics[width=\textwidth]{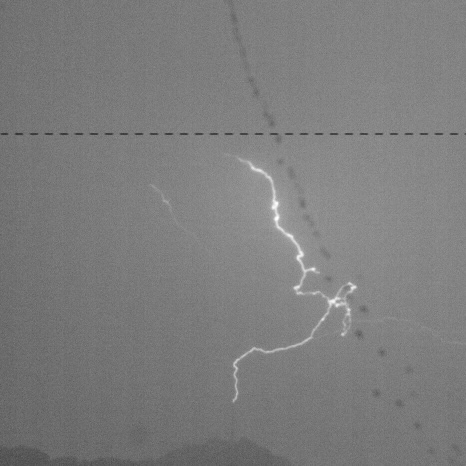}
        \includegraphics[width=\textwidth]{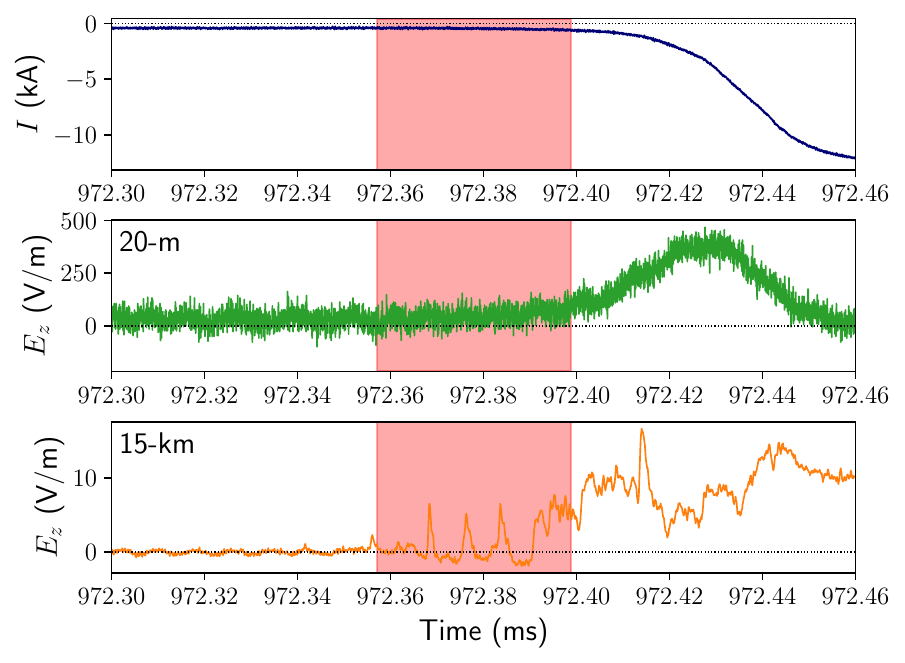}
        \subcaption*{Frame 975}
    \end{subfigure}
    \begin{subfigure}{0.32\textwidth}
        \includegraphics[width=\textwidth]{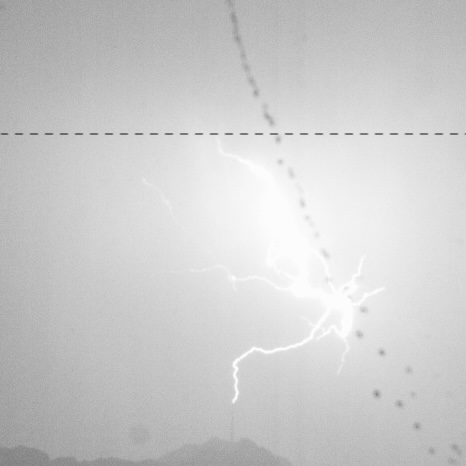}
        \includegraphics[width=\textwidth]{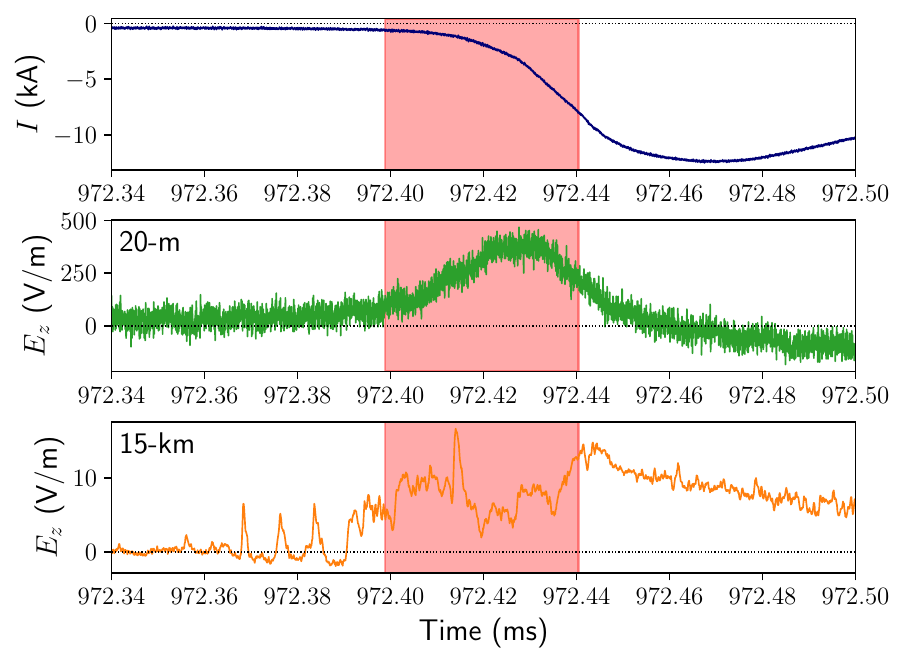}
        \subcaption*{Frame 976}
    \end{subfigure}
    \caption{The lead-up to the main pulse of the flash, which occurred 13.45~ms after the onset of the ICC. The six HSC frames leading up to the arrival of the recoil leader at the junction point are shown above. Yellow arrows highlight the direction of propagation, and a dashed, horizontal line has been added to identify the approximate cloud base ($\sim$900 m above the tower).
    The waveforms are as described in Figure~\ref{fig:UNL}. Note that the background current (estimated by averaging over 400~$\mu$s prior to the main pulse) is approximately $-0.39\pm0.05$~kA.}
    \label{fig:RL-MP}
\end{figure}


\citeA{mazur_physical_2002} defined recoil leaders as negative leaders traversing preexisting positive channels.
In 2019, \citeA{qie_propagation_2019} observed for the first time the inverse case: positive recoil leaders with a positive leader tip retrograding in a preexisting negative channel. 
This is also the case for the presented flash in our study, as observed in Figure~\ref{fig:RL-MP}, which provides a close-up of phase (c).
These high-speed camera frames confirm that the main pulse of this particular upward positive flash was triggered by a downward-connecting recoil leader with a 2D junction height of 370$\pm$5~m above the tower tip (see Figure~\ref{fig:wholeflash}), well below the 1~km cutoff suggested by \citeA{zhou_study_2015} for pulses to be classified as mixed-mode pulses.
The tower current and near E-field waveforms, though expectedly non-zero, display very little change until the connection time, at which point a large current wave propagates from the junction point (Frame 976).
Note that the fast pulses leading up to the main pulse in the 15-km Herisau E-field waveform (Frames 972, 974 \& 975) may indeed be due to the recoil leader, implying a stepped nature. 
Their absence in the Radome (near) field can be explained by the \citeA{azadifar_new_2019} model for M-components in upward flashes: the Distance Range Ratio (DRR) they define ``as the ratio of the peak amplitudes of the microsecond‐scale pulse at the start of the M-component and of the ensuing millisecond‐scale pulse'', was shown to be directly proportional to distance, such that these fast ($\mu$s-scale) pulses associated with the branch current wave vanish in the near field.
Physically-speaking, this is due to their being overwhelmed by the static component of the electric field, which is negligible at a distance of 15 km.
The distinction here is that the $\mu$s-scale pulse in the \citeA{azadifar_new_2019} model is due to a current wave ascending the (recoil or newly-formed leader) branch \textit{post}-connection, while our observed fast pulse\textbf{s}, \textit{prior} to the connection time, are associated with the recoil leader itself, either the positive end descending the branch, or more likely the negative end ascending if it is bidirectional.



%


Interestingly, while the observed main pulse's 12~kA peak current is reasonable for a mixed-mode pulse \cite{he_analysis_2018, he_characteristics_2020}, the current and E-field rise times are both $>$10~$\mu$s, more characteristic of an M-component–type ICC pulse \cite{flache_initialstage_2008}, see Frame 976 of Figure~\ref{fig:RL-MP}.
[Note that higher amplitude currents are expected in mixed-mode compared to M-component--type pulses due to the lower junction point.]
This may be due to signal propagation along the plasma channel (2D length of $>$730~m from the tower tip to the junction point), as opposed to through the air ($\sim 425$~m), which results in an increase in the current rise time, and subsequently the E-field rise time.
These lengths were estimated from the Kr\"onberg HSC (as the Schw\"agalp HSC’s field of view was limited to the bottom of the channel); dividing the former by the time it took for the upward negative leader to reach the junction point, the average 2D velocity of the upward negative leader was estimated to be $\sim 9 \times 10^4$~m/s. 
This is on the lower end of the range of 3D velocities during horizontal propagation established by \citeA{wu_upward_2020}:
0.9 to $7.3 \times 10^5$~m/s with a mean value of $3.8 \times 10^5$~m/s. 
Although the Schw\"agalp HSC’s sensitivity was set to capture return strokes, too low to observe the upward negative leader-stepping phase of the flash depicted in Figure~\ref{fig:UNL}, we were able to reconstruct the 3D geometry of the bottom of the channel from overlapping fields-of-view using later-time frames 
(see Video in the Supplementary Information), and subsequently estimate the 3D leader velocity to be $\sim 2.4 \times 10^5$~m/s 
from the early-time Kr\"onberg HSC frames, such as those presented in Figure~\ref{fig:UNL}.
This aligns much better with the 3D speeds of \citeA{wu_upward_2020}, both for horizontal propagation (discussed above), as well as vertical propagation: 1.8 to $27.9 \times 10^5$~m/s, albeit lower than their mean value of $10.4 \times 10^5$~m/s. 
In turn, we can multiply our new velocity by the time ($\sim$8.38~ms) between the onset of the ICC (Figure~\ref{fig:UNL}) and the creation of the future recoil leader's branch, depicted in figures~\ref{fig:branch} \& \ref{fig:branchHSC}, to more accurately estimate the 3D tower-tip-to-junction-point channel length as $\sim$2010~m. 

Analysis of frames 972 through 975 in Figure~\ref{fig:RL-MP} can approximate the downward propagation velocity (2D average) of the recoil leader to be $\sim 4 \times 10^6$~m/s. 
This was done by measuring the 2D length of that channel branch in pixels ($\sim$200, as estimated from Frame 975, when it is fully illuminated), converting to meters ($\sim670\pm5$) using the Tower as a reference (124~m / 37~pix), and dividing by the time (167~$\mu$s) associated with the 4 frames between the appearance of the recoil leader and the onset of the Main Pulse. 
[The 1-2~pix uncertainty similarly converts to a margin of error at most 10~m, and subsequently a speed uncertainty of $\pm 3 \times 10^4$~m/s.]
Even as a lower bound on the true value, this is already over an order of magnitude faster than most previous estimates of positive leader speeds. 
\citeA{biagi_observations_2010} and \citeA{jiang_high-speed_2014} calculated the 2D velocities of rocket-triggered upward positive leaders to be in the range of $0.55 – 1.8 \times 10^5$~m/s, while an analysis by \citeA{wu_upward_2020} of negative cloud-to-ground and intra-cloud flashes found positive leader velocities around $1 - 3 \times 10^4$~m/s.
\citeA{montanya_start_2015} observed a ``virgin air'' bidirectional leader, and determined its positive end to propagate with a (decreasing) speed of $1 - 9 \times 10^4$~m/s.
Our observed 2D partial speed was comparable, however, to that measured by \citeA{qie_propagation_2019} for the positive end of a bidirectional recoil leader in an upward positive flash of Type 2 (as defined by \citeA{romero_positive_2013}; see next section), at $6.4 \times 10^6$~m/s.
In their study, the observed positive tip propagated upwards into the cloud, while the negative tip propagated downwards, eventually connecting to the main (upward negative) leader channel.
Considering our upward positive flash was of Type 1, and it was the \textit{positive} leader tip that connected to the main leader channel, we propose this unique configuration of charge transfer as the triggering mechanism for the characteristic main pulse.

Previous studies have demonstrated the impact of preceding nearby lightning activity on the formation of upward leaders \cite{warner_observations_2012, smorgonskiy_model_2015, wu_upward_2020, yuan_cloud_2021, sunjerga_initiation_2021}; it is therefore worth noting that this flash saw no activity within a 30 km radius in the 3 seconds prior to initiation (as confirmed from the near E-field record and the EUCLID Lightning Location System), and can therefore be classified as ``self-triggered''. Consistently, our electric field records show no activity during the 960 ms preceding the initiation of the upward flash.

It should also be noted that this flash occurred during the Laser Lightning Rod project presented in \citeA{houard_laser-guided_2023} (therein labelled L2), while the laser was on. The guiding effect discussed therein was confined to early-stage propagation, only observed over a distance of about 50~m (see their Fig. 2), and therefore should not have impacted the effects studied in this article, which occurred an order of magnitude higher. See \citeA{oregel-chaumont_direct_2024} for further discussion.


\section{Discussion}
\label{sec:dis}

\citeA{romero_positive_2013} classified upward positive flashes into two categories: Type 1 flashes, which exhibit a large unipolar return stroke–like current pulse following the upward negative stepped-leader phase, and Type 2 flashes, which do not feature such a large pulse and consist solely of a 100 millisecond-scale waveform with large, oscillatory pulse trains due to upward negative stepped leaders.
From the waveforms presented in figures~\ref{fig:wholeflash} \& \ref{fig:RL-MP}, our flash was clearly a Type 1 upward positive flash; the characteristics of its Main Pulse are summarized in Table~\ref{tab:mpchar}: (from left to right) the time from the onset of the stepped leader, the junction height and length, the absolute value peak current, the current pulse 10-90\% rise time, the associated near and far electric field changes and rise times, the time lag between the current and the 20-m electric field, and the asymmetrical waveform coefficient of the current pulse, as defined in \citeA{he_analysis_2018}, see below.
We also estimated the charge transferred during the Main Pulse to be $\sim 2.3$~C, a value consistent with previous observations \cite{he_characteristics_2020}. 


\begin{table}
\centering 
\caption{Main pulse characteristics}
\begin{tabular}{ c | c | c | c | c | c | c | c | c | c | c } %
\hline 
$t_{SL}$ & \multicolumn{2}{c}{Junction [m]} & $I_p$ & $t_{10-90}$ & \multicolumn{2}{c}{$\Delta E$ [$\frac{\mathrm{V}}{\mathrm{m}}$]} & \multicolumn{2}{c}{$t_{Er}$ [$\mu$s]} & $t_\mathrm{lag}$ & AsWC \\ 
{[ms]} & 2D Height & 3D Length & [kA] & [$\mu$s] & 20 m & 15 km & 20 m & 15 km & [$\mu$s] & \\
\hline 
13.45 & 370$\pm$5 & $\sim$2010 & 12.10$\pm$0.05 & 34.2 & 470$\pm$40 & 16.4$\pm$0.3 & $\sim$40 & $\sim$20 & $\sim$40 & 0.83 \\
\hline 
\end{tabular}
\label{tab:mpchar}
\end{table}

\begin{figure}
    \begin{subfigure}{\textwidth}
        \includegraphics[width=\textwidth]{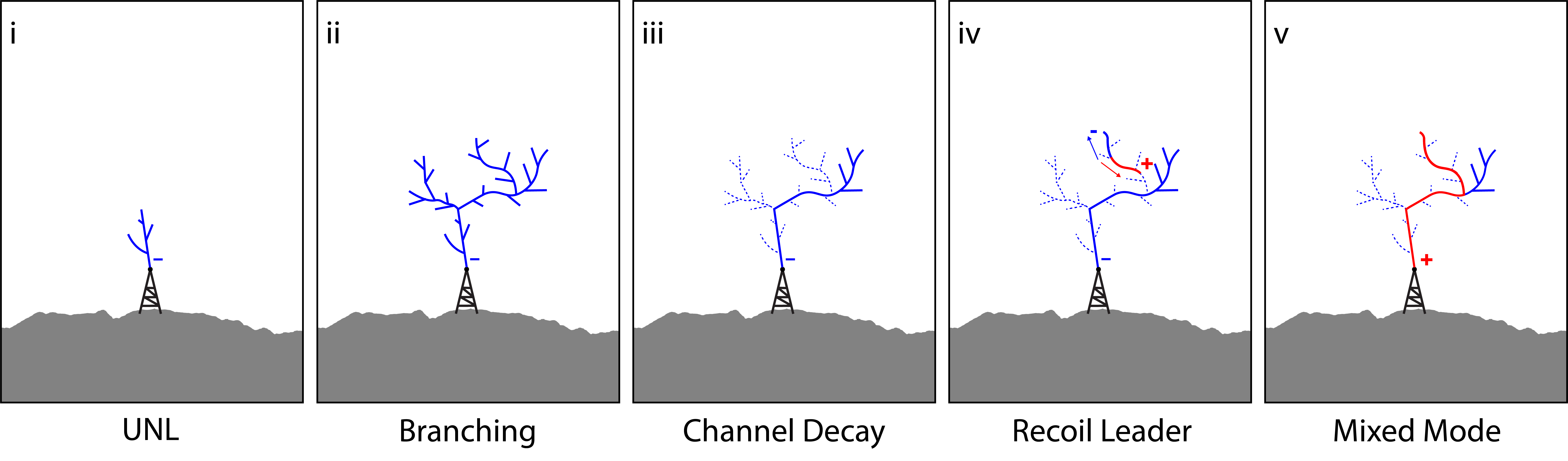}
        \subcaption{Type 1 flashes}        
    \end{subfigure}
    \begin{subfigure}{\textwidth}
        \includegraphics[width=.6\textwidth]{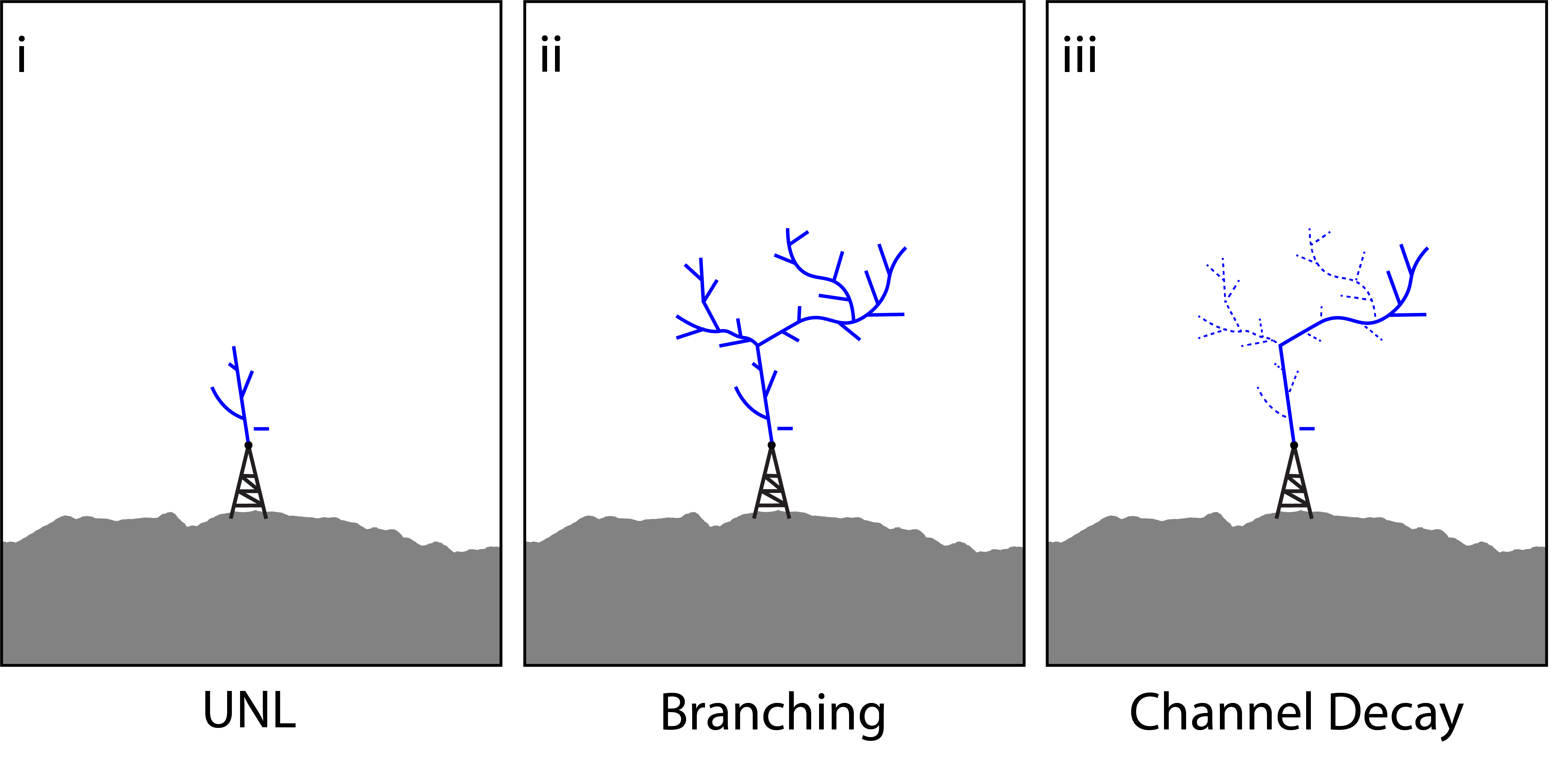}
        \subcaption{Type 2 flashes}        
    \end{subfigure}
    \caption{Diagram depicting the underlying mechanisms in a typical upward positive flash. Red and blue indicate positive and negative charge transfer, respectively, while solid and dashed lines indicate active and decayed branches, respectively. The top set of panels (a) show the underlying mechanisms in Type 1 flashes where the main pulse is due to the connection of a positive recoil leader tip with the main branch, as illustrated in (iv) and (v). Type 2 flashes, depicted in the bottom set of panels (b), end after the third phase (iii), i.e., the current gradually dies out as the plasma channels decay.} 
    \label{fig:UPFtypes}
\end{figure}

The classification scheme of \citeA{romero_positive_2013} was based exclusively on current waveform observations. 
The analysis presented in the present paper allows us to propose a physical mechanism describing Type 1 and Type 2 positive flashes, as illustrated in Figure~\ref{fig:UPFtypes}. 
Both types begin with (i) an upward negative leader (UNL) initiated at the tip of the strike object, which undergoes significant branching (ii) as it rises towards the cloud. 
Most of these branches decay (iii), leaving a main, current-carrying channel. 
This is where it ends for Type 2 upward positive flashes. 
Type 1 flashes, however, see a (possibly stepped and bidirectional) recoil leader descending a defunct plasma channel (iv), to reconnect with the main branch, triggering their characteristic main pulse (v) which propagates from the junction point.


%


The high-speed camera frames in Figure~\ref{fig:branch} presented earlier-time evidence of the existence of the plasma channel later retraced by the recoil leader.
It can be seen from the waveforms in Figure~\ref{fig:RL-MP} that this recoil leader (possibly bidirectional, though the ascending negative tip is not visible if so) initiated a mixed-mode charge transfer of positive polarity, as inferred from the signs of the current pulse and associated E-field changes. 
It is of interesting note that positive recoil leaders should play a role in upward positive flashes, in a similar way as \citeA{sunjerga_bidirectional_2021} demonstrated that \textit{negative} recoil leaders play a role in upward \textit{negative} flashes.


\begin{table}[t]
\centering
\caption{Existing criteria for distinguishing MM and M-C type ICC pulses}
\begin{tabular}{ c | c | c | c | c }
\hline
Type & $t_{10-90}$ [$\mu$s] & $t_\mathrm{lag}$ [$\mu$s] & Junction height [km] & AsWC \\ 
 & \protect\cite{flache_initialstage_2008} & \protect\cite{zhou_study_2015} & \protect\cite{zhou_study_2015} & \protect\cite{he_analysis_2018} \\ \hline
MM  & $<8$ & $<10$ & $<1$ & $>0.8$ \\
M-C & $>8$ & $>10$ & $>1$ & $<0.8$ \\ 
\hline
\end{tabular}
\label{tab:iccpchar}
\end{table}


As discussed in Section~\ref{sec:intro}, \citeA{zhou_study_2015} also used the time interval between the onset of the current pulse and the peak of the electric field to distinguish between M-component type and mixed-mode pulses, which is equivalent to a height criterion of the attachment point of the rejuvenated branch attaching to the main channel.
Table~\ref{tab:iccpchar} presents the distinction criteria proposed in the literature: `MM' and `M-C' stand for mixed-mode and M-component -type ICC pulses, respectively; 
$t_{10-90}$ is the 10--90\% current rise time in microseconds \cite{flache_initialstage_2008};
$t_\mathrm{lag}$ 
the time lag between the current and E-field peaks in microseconds \cite{zhou_study_2015} [Note that this value depends on the E-field measurement distance; \citeA{zhou_study_2015} based their 10~$\mu$s cutoff on the \citeA{rakov_mechanism_1995} 30~m, whereas our sensor is at 20~m. This is an acceptable difference as the junction point is over an order of magnitude further away.];
and AsWC the asymmetrical waveform coefficient of the current pulse, defined by \citeA{he_analysis_2018} as:
\begin{equation}
    \mathrm{AsWC} = \frac{\mathrm{FWHM} - t_{50-100}}{\mathrm{FWHM}} 
\end{equation}
where FWHM 
and $t_{50-100}$ 
are the full width at half maximum and 50--100\% rise time, respectively.
Note that the criteria presented by \citeA{zhou_study_2015} are based on the implicit assumption of a vertical channel.
Since the main plasma channel of the flash discussed here exhibited significant tortuosity with a substantial horizontal component, the tower-tip-to-junction-point length may be a better criterion than the junction height for capturing the effects of channel geometry. 
Furthermore, it is worth noting that $t_\mathrm{lag}$ depends on the distance at which the electric field is measured. 
As a result, the time interval criterion should ideally be adjusted as a function of the measurement distance. 
In this context, we believe that the junction point is a more appropriate criterion for distinguishing between mixed-mode and M-component--type pulses.
To this point, Table~\ref{tab:mpcrit} shows how the characteristics of the observed flash's main pulse are such that it could be classified as either a mixed-mode or M-component--type ICC pulse, depending on which criterion is considered.
Specifically, the 2D junction height of 370~m and AsWC $\approx$ 0.83 would classify it a mixed mode [note the proximity of our calculated AsWC to the defined 0.8 cutoff],
while the 3D channel length of 2~km, the $t_{10-90} \approx 34$ $\mu$s, and the $t_\mathrm{lag} \approx 40$ $\mu$s, all seem to suggest an M-component--type.


\begin{table}%
\centering
\caption{Classification of the main pulse as either Mixed Mode (MM) or M-component--type ICC pulse, based on the existing criteria}
\begin{tabular}{ c | c | c | c | c | c }
\hline
\multicolumn{2}{c}{Junction} & $I_p$ & $t_{10-90}$ & $t_\mathrm{lag}$ & AsWC \\
2D Height & 3D Length &  &  & \\ \hline
MM & M-C & MM & M-C & M-C & MM \\
\hline
\end{tabular}
\label{tab:mpcrit}
\end{table}


As discussed in Section~\ref{sec:intro}, since both modes involve the continuous current mode of charge transfer in addition to a second mode of charge transfer (M-component--type or return stroke-like), the term ``mixed mode'' applied to only one type of pulses may not be the optimal choice as further discussed below.
If we assume that branches can indeed attach to the channel at any height, then a gradual transition between these wave shapes should be observable. 
The terms M-component type and mixed mode, applied to cases in which the junction point is, respectively, far or close to the channel base, may therefore not fully capture the spectrum of behaviors observed, as one would expect a continuum of pulse rise times and symmetries that vary with the junction height (or the length of the channel to the attachment point).




\section{Conclusions}

Herein we presented simultaneously recorded current, electric field and high-speed video footage of an upward positive flash initiated from the S\"antis Tower in Switzerland. The analysis provided a physical mechanism distinguishing Type 1 and Type 2 upward positive flashes. 

We have shown that the Type 1 main current pulse can be the result of a recoil leader-initiated mixed mode of charge transfer, determined by the current and electric field records to be of positive polarity, the first inference of its kind.

The estimated propagation speed of this recoil leader’s positive tip was consistent with prior measurements, but unique in being the first observation in a Type 1 upward positive flash.

Furthermore, the recoil leader was accompanied by a series of fast electric field pulses that suggest step-like propagation: to our knowledge another observational first.
More data are needed to confirm that the proposed mechanism is the only one describing a Type 1 flash.

We also discussed the ambiguity in distinguishing mixed mode and M-component--type pulses, according to existing criteria based on pulse characteristics and channel geometry. 
As discussed by \citeA{zhou_study_2015}, the key phenomenological difference between these two phenomena is the height of the junction point, which also determines the time lag between the electric field and the current. 
Specifically, the terms M-component--type and mixed mode refer to cases in which the junction point is, respectively, far from (higher than 1 km) or close to (lower than 1 km) the channel base. 
This definition may not fully capture the spectrum of the observed behaviors, as one would expect a continuum of pulse characteristics (such as rise times and symmetries) that vary with the position of the junction point along the channel. 
This ambiguity can be resolved by considering the two phenomena as extreme cases of a single process involving the attachment of a downward leader or recoil leader to the channel.

These observations contribute to improving our understanding of the mechanisms of charge transfer in upward lightning flashes, and will be furthered by in-depth analysis of differences between the various types.

\newpage

\appendix
\section{Branching HSC Frames}


\begin{figure}
    \centering
    \begin{subfigure}{0.24\textwidth}
        \includegraphics[width=\textwidth]{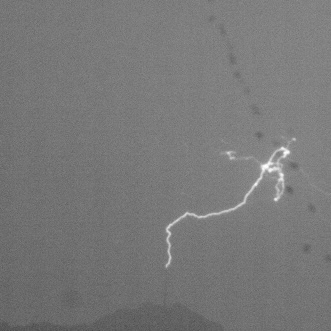}
        \subcaption*{Frame 864}
    \end{subfigure}
    \begin{subfigure}{0.24\textwidth}
        \includegraphics[width=\textwidth]{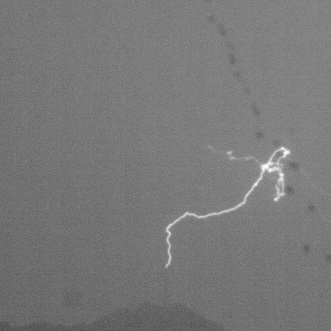}
        \subcaption*{Frame 865}
    \end{subfigure}
    \begin{subfigure}{0.24\textwidth}
        \includegraphics[width=\textwidth]{images/cropped_frame_866.jpg}
        \subcaption*{Frame 866}
    \end{subfigure}
    \begin{subfigure}{0.24\textwidth}
        \includegraphics[width=\textwidth]{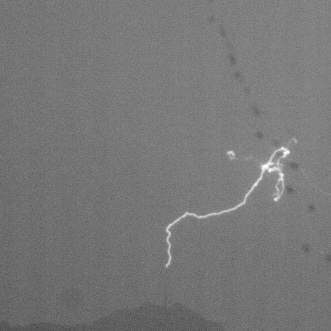}
        \subcaption*{Frame 867}
    \end{subfigure}
    \begin{subfigure}{0.24\textwidth}
        \includegraphics[width=\textwidth]{images/cropped_yellip_frame_868.jpg}
        \subcaption*{Frame 868}
    \end{subfigure}
    \begin{subfigure}{0.24\textwidth}
        \includegraphics[width=\textwidth]{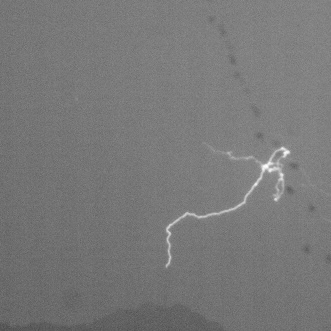}
        \subcaption*{Frame 869}
    \end{subfigure}
    \begin{subfigure}{0.24\textwidth}
        \includegraphics[width=\textwidth]{images/cropped_yellip_frame_870.jpg}
        \subcaption*{Frame 870}
    \end{subfigure}
    \begin{subfigure}{0.24\textwidth}
        \includegraphics[width=\textwidth]{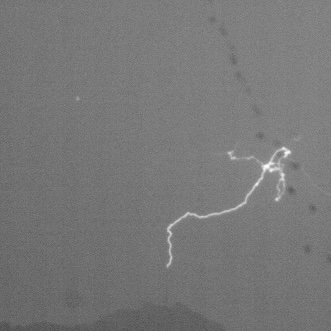}
        \subcaption*{Frame 871}
    \end{subfigure}
    \begin{subfigure}{0.24\textwidth}
        \includegraphics[width=\textwidth]{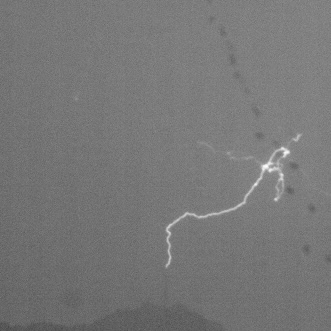}
        \subcaption*{Frame 872}
    \end{subfigure}
    \begin{subfigure}{0.24\textwidth}
        \includegraphics[width=\textwidth]{images/cropped_yellip_frame_873.jpg}
        \subcaption*{Frame 873}
    \end{subfigure}
    \begin{subfigure}{0.24\textwidth}
        \includegraphics[width=\textwidth]{images/cropped_yellip_frame_874.jpg}
        \subcaption*{Frame 874}
    \end{subfigure}
    \begin{subfigure}{0.24\textwidth}
        \includegraphics[width=\textwidth]{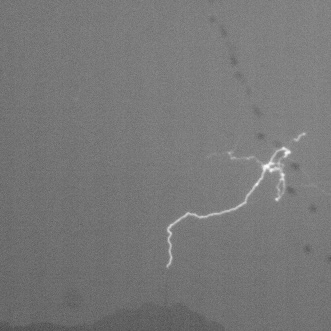}
        \subcaption*{Frame 875}
    \end{subfigure}
    \begin{subfigure}{0.24\textwidth}
        \includegraphics[width=\textwidth]{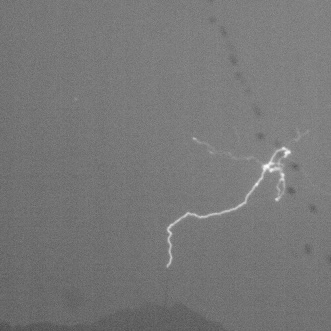}
        \subcaption*{Frame 876}
    \end{subfigure}
    \begin{subfigure}{0.24\textwidth}
        \includegraphics[width=\textwidth]{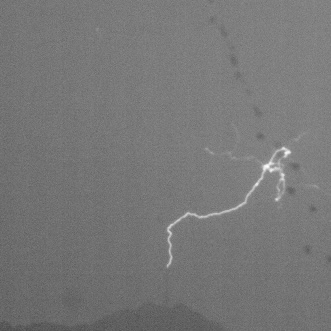}
        \subcaption*{Frame 877}
    \end{subfigure}
    \begin{subfigure}{0.24\textwidth}
        \includegraphics[width=\textwidth]{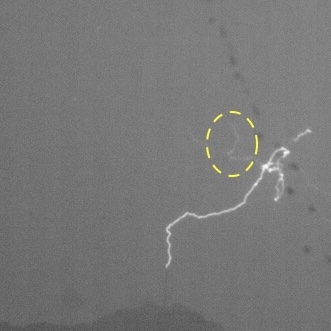}
        \subcaption*{Frame 878}
    \end{subfigure}
    \begin{subfigure}{0.24\textwidth}
        \includegraphics[width=\textwidth]{images/cropped_yellip_frame_879.jpg}
        \subcaption*{Frame 879}
    \end{subfigure}
    \caption{HSC frames showing the creation of the plasma channel (2D junction height of 495$\pm$5 m AGL) along which the recoil leader will later propagate. The branch first appears noticeably in Frame 868, and further demonstrates significant propagation in Frames 870, 873, 874, 878, and 879 (highlighted in yellow). The time between two consecutive frames is $\sim$42~$\mu$s.}
    \label{fig:branchHSC}
\end{figure}


\newpage

%
%

%

%

\section*{Open Research}

Raw and processed data sets, as well as the custom code used for data processing and analysis in this study, are available on Zenodo \cite{oregel-chaumont_dataset_2024} and GitHub (\url{https://github.com/TomaOC/santis}).





\acknowledgments
This work was supported in part by the Swiss National Science Foundation (Project no.s 200020\_175594 and 200020\_204235) and the European Union's Horizon 2020 research and innovation program (grant agreement no. 737033-LLR).
The authors would like to thank Florent Aviolat for developing a data-visualization software that expedited flash analysis, and Jean-Pierre Wolf for confirming the 3D plasma channel structure, presented as supplementary material, with an independent analysis.


%
%



\bibliography{UPF_Mixed-modes.bib}

%
%
%
%
%

\end{document}